# Impedance Measurements of the Extraction Kicker System for the Rapid Cycling Synchrotron of China Spallation Neutron Source*

Liangsheng Huang[1,2], Sheng Wang[1,2], Yudong Liu[1,2,#], Yong Li[1,2], Renhong Liu[3], Ouzheng Xiao[3]

[1] Dongguan Campus, Institute of High Energy Physics (IHEP), Chinese Academy of Sciences (CAS), Dongguan, 523803, China

[2] Dongguan Institute of Neutron Science (DINS), Dongguan, 523808, China

[3] Institute of High Energy Physics, Chinese Academy of Science, Beijing, 100049, China

**ABSTRACT**

The Rapid Cycling Synchrotron (RCS) of China Spallation Neutron Source (CSNS) includes eight modules of the ferrite kicker magnets with window-frame geometry. The fast extraction kicker system is one of the most important accelerator components, whose inner structure will be the main source of the impedance in the RCS. It is necessary to understand the kicker impedance before its installation into the tunnel. The conventional and improved wire methods are employed for the benchmarking impedance measurement results. The experimental result on the kicker's impedance is explained by comparison with the CST PARTICLE STUDIO simulation. The simulation and measurements confirm that the window-frame ferrite geometry and the end plate are the important structures causing the coupling impedance. It is proved in the measurements that the mismatching among the long cable, the termination and the power form network lead a serious oscillation sideband in the longitudinal and vertical impedance, which can be restricted by the ferrite absorb ring. The total impedance of the eight modules systems is determined by the scaling law from the measurement and the impedance measurement of the kicker system is summarized.

＊ Supported by National Natural Science Foundation of China (11175193, 11275221)

# liuyd@ihep.ac.cn

1. Introduction

   The driving terms of beam instabilities in accelerator depend on the interaction between the charged particles and the surroundings which are usually described by the coupling impedance. The longitudinal coupling impedance may lead to beam energy loss, thus heating of components and energy spread [1]. The transverse impedance attributes to the instability when the off-axis beam passing through vacuum components. For every new particle accelerator, careful establishment of an "impedance budget" is a prerequisite for achieving expected performance. Therefore, theoretical analyses, simulation and measurement on the bench of the coupling impedance are crucial tasks in the accelerator design, development, and research. The analytical formulas of some vacuum components are given in references [2][3][4], and some developed code can be used to simulate the coupling impedance, such as ABCI, HFSS [5] and CST PARTICLE STUDIO [6], but the coupling impedance of very complicate components is difficult to study through the analytical formula and the simulation code, so the impedance measurement on the bench is useful at that time.

   The conventional wire-method has been widely employed in the coupling impedance measurement on the bench, which are insert on-axis wire (coaxial-wire method) into the Device Under Test (DUT) to measure the longitudinal impedance and insert two parallel wires with out of phase wire current (twin-wire method) into the DUT to measure the transverse impedance. The longitudinal coupling impedance measurement with the coaxial wire method was firstly achieved by A. Faltens in 1971 [7]. M. Sands and J. Rees exhibited the coupling impedance measurement in the time domain [8], but the impedance measurement in the frequency domain is widely used as the development of technique. G. Nassibian measured the transverse impedance in 1978 by the loop method [9], the origin of the twin-wire method. Mr Walling measured the transverse impedance by the twin-wire method in 1987 [10] and he gave the famous formula of the distributed component impedance - the logarithmic formula. F. Caspers and H. Hahn gave some useful and comprehensive document for the impedance bench measurement [11][12]. Moreover, the validity of the measured longitudinal impedance by the coaxial-wire has been interpreted by H. Hahn [13]. There has been a long history in the development of the wire method and the improvement of its accuracy in both theory and technique, the wire method, therefore, is called standard method for the impedance measurement. In addition, the twin-wire method is developed recently since the longitudinal coupling impedance measurement by the twin-wire method with in phase wire current (common-mode signal) is reported [14], so it is convenient for the twin-wire method to measure the longitudinal and transverse impedance together. The measured method and example of the impedance is shown in many papers [15]-[20].

The basic principle of the impedance measurement with the wire method is to use the wire current to simulate the beam current. In this equivalent, there are two approximations [8]: first, there must be similarity on electromagnetic field produced by the wire current and the beam current. Second, the distortion on the wire current can represent the energy variety of the beam current. Additionally, the algebraic expression of the impedance should be decided for the bench measurements. The error of the bench measurement, therefore, comes from the wire current approximation and the approximate expression of the impedance measurement.

The China Spallation Neutron Source (CSNS) accelerators consist of an $H^-$ linac, a proton Rapid Cycling Synchrotron (RCS), and two beam transport lines [21]. The RCS is designed to accelerate the proton beam from 80 MeV to 1.6 GeV with the repetition rate of 25 Hz. The beam power of the CSNS accelerator is 100 kW, and it will be upgraded to 500 kW. Due to the high beam intensity and high repetition rate, the ratio of beam loss must be controlled to a very low level. In the case of the RCS, the extraction kicker represents the most critical impedance item [22], so it should be measured on the bench. The bunch length in the RCS is shortened from 420 ns to 80 ns. The frequency range of interest to the CSNS covers frequencies, from ~100 MHz down to below 1 MHz, with emphasis on the 1 - 12 MHz range. The coaxial-wire method is used to measure longitudinal coupling impedance. The transverse coupling impedance is firstly measured by the twin-wire method. Applying the twin-wire method to the low frequency part entails difficulties [23], especially below 10 MHz, the signals become very weak and the noise and drift of the instrument lead to results which may no longer be valid, thus, one turn loop is been adopted to measure the transverse impedance on low frequency, and the input impedance is measured.

The extract kicker system is very complicate, the Pulse Form Network (PFN), the 12.5 Ω termination and the connection long cable contribute to the longitudinal and vertical impedance, so the kicker system is firstly introduced in the section 2. The longitudinal and transverse coupling impedances measurements are exhibited in section 3 and section 4, respectively. Section 5 is the summary of the kicker impedance study.

## 2. The extraction kicker system of CSNS/RCS

There are eight kickers in CSNS/RCS [24][25], a total kick strength of 20 mrad for the kickers will create about 130 mm central orbit displacement at the entrance of the lambertson magnet, which is required by the separation of the extracted beam, the cycling beam and the septum thickness. The parameters of the kickers are shown in Table 1. The schematic view of a kicker system is shown in Fig 1, and the main part of the kicker system is the window-frame geometry in Fig 2 and its material is CMD5005 ferrite. The side strap at the side of the magnet in Fig 1 connects the upper and lower busbar plates, the busbar and window-frame ferrite is located in a vacuum vessel, 0.58 m length. The busbar is fully isolated from the vessel by the ceramic block of the feedthru. The eddy current strip (ECS) located in vertical center of the ferrite block to reduce the longitudinal impedance is invisible in Fig 2. The total inductance of a kicker is about 0.9 μH. The capacitance is about 30 pF mainly from the end plate in Fig 1.

The power of the magnet comes from the PFN [26] via the end plate, the feedthru and the 130 m length cable. The characteristic impedance of the cable is 12.5 Ω, and the impedance of the PFN is 6.25 Ω. The PFN matches the cable which parallels the

12.5 Ω termination during the time of the beam extraction. The saturated reactor is designed to cut the PFN when the beam is accelerated, at that time, the cable connects on the termination, and it matches. Due to the error of the cable and the distributed effect of the termination [17][27], the cable, the termination and the PFN mismatch actually, which affects the longitudinal and vertical impedance. Therefore, the measurement divides two parts: the kicker without the long cable, the PFN and the termination (naked kicker) and the kicker connects the cable, the PFN and the termination on (kicker system).

The No. 2 kicker (K2) is firstly manufactured. Based on the kicker, the longitudinal and transverse impedances are measured.

Table 1: The parameters of the eight kickers

| Number | kicker 1 | Kicker 2 | Kicker 3 - 4 | Kicker 5 - 6 | Kicker 7 - 8 |
| --- | --- | --- | --- | --- | --- |
| Strength (T) | 0.0558 | 0.0559 | 0.0526 | 0.0571 | 0.0609 |
| Angle (mrad) | 2.6 | 2.2 | 2.99 | 2.9 | 2.76 |
| Length (m) | 0.4 | 0.32 | 0.45 | 0.41 | 0.36 |
| Top time (ns) | >550 | >550 | >550 | >550 | >550 |
| Width (mm) | 155 | 212 | 155 | 157 | 168 |
| Gap (mm) | 136 | 144 | 153 | 141 | 132 |
| Rise time (ns) | ⩽250 | ⩽250 | ⩽250 | ⩽250 | ⩽250 |

Fig 1: The schematic view of the CSNS/RCS kicker system

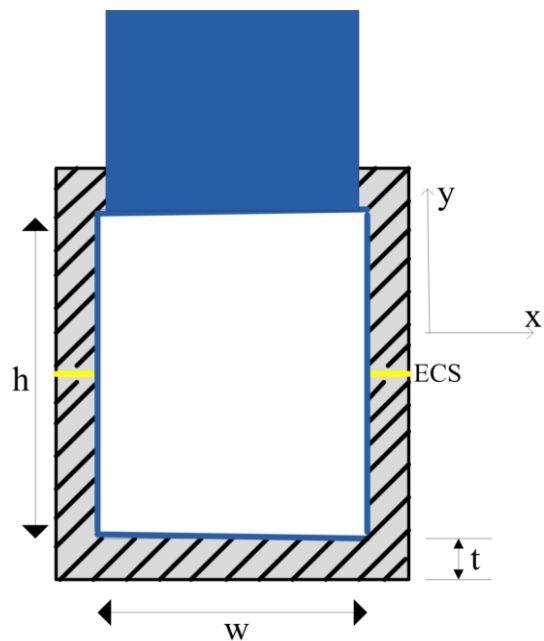

Fig 2: The schematic view of the window-frame geometry

## 3. The longitudinal impedance measurement

The schematic setup of the coaxial-wire method is shown in Fig 3. The kicker with a thin copper wire on its beam axis can be regards as a two-port microwave circuit, of which the forward scatter coefficient can be measured with the Vector Network Analyzer (VNA). The measurement constantly requires two independent, consecutive measurements of the forward scatter coefficient $S_{21}$ of the DUT and a smooth reference beam pipe (REF) of equal length. The longitudinal coupling impedance can be found [10] from the measured transmission coefficients $S_{21}$ as

$$Z = -2R_c \ln(\frac{S_{21,DUT}}{S_{21,REF}}), \tag{1}$$

here, $S_{21,\,DUT}$ and $S_{21,\,REF}$ are the forward scatter coefficient of the DUT and the REF, respectively. $R_c$ is characteristic impedance formed by the reference pipe with the copper wire as a coaxial transmission line structure. The radius of beam pipe and the wire are 125 mm and 0.25 mm, respectively. Matching the system impedance of the VNA (50 Ω) to $R_c$ of the reference line (372 Ω) is achieved by adding a series metal resistor ($R_s$, 320 Ω) at the each end of the line. The matching resistor is shielded by the 35 mm SUCOBOX [28] on the right. The type of the network analyser - Agilent E5071C is used in the measurement. In order to decrease the temperature drift, the temperature is stable at 25 degrees in the experiment.

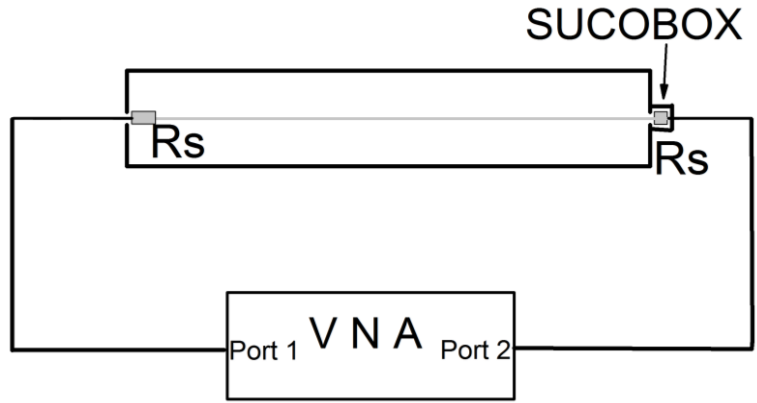

Fig 3: The schematic setup of the longitudinal coupling impedance measurement

The forward transmission coefficients for the REF, the naked kicker and the kicker system are measured. Replacing the coefficient into Eq. (1), the longitudinal coupling impedance of the naked kicker and the kicker system are obtained and shown in Fig 4 and Fig 5. Two peaks on about 18 MHz and 30 MHz in Fig 4 come from the end plate and the window-frame ferrite geometry, respectively. The contribution of the end plate is exported by the 130 meters length cable, so the peak on 18 MHz disappears in Fig 5 when the cable, the PFN and the termination connect on the naked kicker, but it is clear to see that an oscillation appears at that time. The space of the oscillation is 0.72 MHz. The space of the reflection from the long cable can be expressed theoretically as

$$\Delta f = \frac{c}{2\alpha L}, \qquad (2)$$

here, $L$ is the length of the cable. The speed factor of the cable medium, $\alpha$, is 1.6 for the polythene. The space is 0.72 MHz, which is absolutely consistent with the result of the measurement. Therefore, it is roughly certain that the oscillation comes from the mismatch among the long cable, the termination and the power form network. It matches perfectly on low frequency and the oscillation sideband is invisible, which is the characteristic of distribution effect of the termination.

To improve the validity of the measured longitudinal impedance, the kicker impedance is simulated by CST PARTICLE STUDIO [6]. Due to excessive memories and CPU time, it is mostly impossible to simulation the real kicker system with the 130 m coaxial cable. Therefore, the impedance of the naked kicker is simulated. A highly simple CST simulation model is constructed, which only includes the window ferrite, the busbar, the end plate, the feedthru and the vacuum tank. Fig 4 gives the comparison on the measurement and the simulation impedances of the naked kicker. The measured longitudinal impedance agrees well with the simulation results.

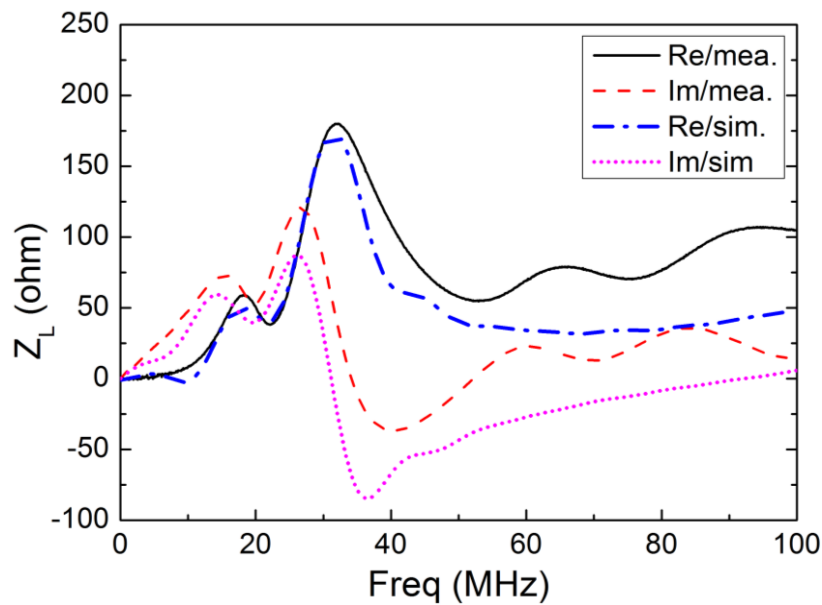

Fig 4: The longitudinal coupling impedance of the naked kicker

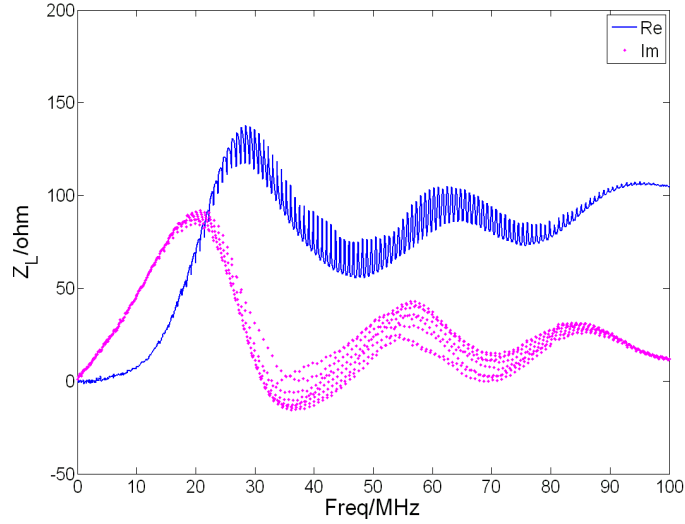

Fig 5: The longitudinal coupling impedance of the kicker system

One of the missions in the measurement is to decrease the impedance, so 8C12 ferrite ring [29][30] is added in the feedthru to restrict the oscillation. The oscillation sideband of the longitudinal impedance mostly disappears in Fig 6 when the ferrite ring is applied. Therefore, the ferrite ring is useful to absorb the reflection.

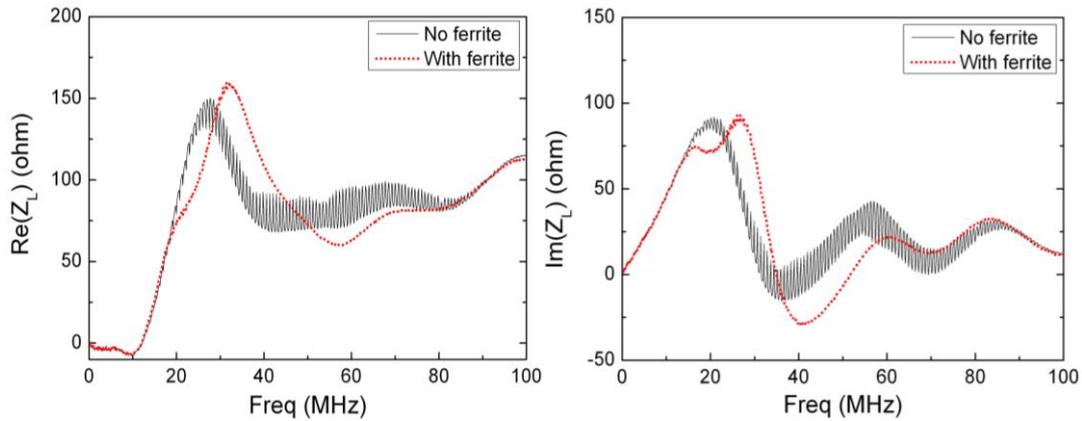

Fig 6: The longitudinal impedance of the kicker system with and without the 8C12 ferrite ring

The entire CSNS/RCS fast extraction kicker system consists of eight individual kickers. The kicker construction of these magnets is similar although different geometrical dimensions in Table 1. Base on CST PARTICLE STUDIO simulation by changing the scale size of the window-frame geometry, the approximate expression of the longitudinal coupling impedance of others kicker is found as

$$Z_i = Z_2 \frac{S_2}{S_i} \frac{L_i}{L_2}, \quad i = 1, 3, \ldots 8. \tag{3}$$

Here, $S_i$ is the area of inner surface of the window-frame geometry of $i^{th}$ kicker, $L_i$ is the length of the window and $Z_i$ is longitudinal impedance of the $i^{th}$ kicker. $S_2$ and $L_2$ are the area of inner surface and the length of the window of the K2 kicker, and $Z_2$ is the measured longitudinal impedance of the K2 in Fig 5.

The longitudinal average impedance of the eight extraction kicker systems is shown in Fig 7, the peak impedance of the extraction kicker system is about 53+$j$45 Ω.

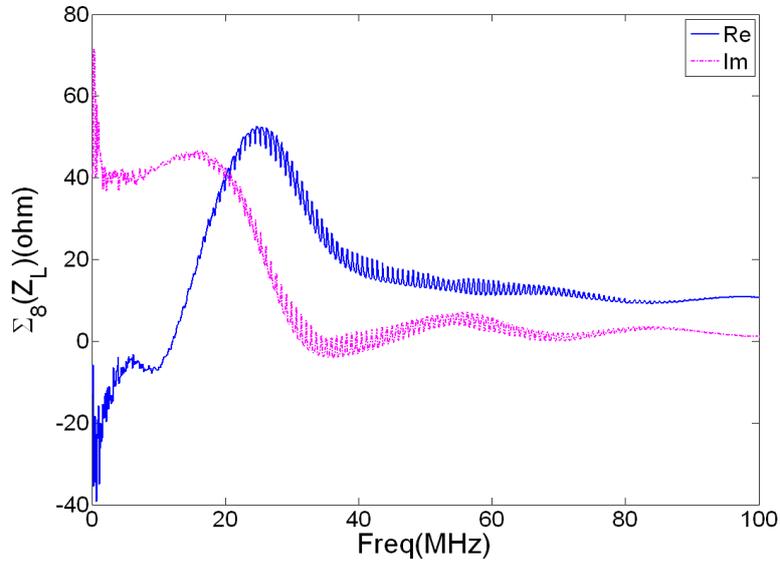

Fig 7: The longitudinal average impedance of the eight kicker systems

### 4. The transverse impedance measurement

For the transverse coupling impedance measurement with the wire method, one standard way insert two parallel wires with out of phase signal (differential-mode) into the DUT in order to produce a dipole current moment, which is called the twin-wire method. The forward scatter coefficient, $S_{21}$, of the twin-wire can be measured by the network analyzer. The schematic view of the transverse coupling impedance measurement is shown in Fig 8. The spacing (2$d$) of the 0.5 mm diameter copper wire is 40 mm, and the characteristic impedance of the twin-wire ($R_c$) is 603 Ω [31], thus the 250 Ω metal resistors are used to match. The differential-mode signal is produced by the MINI hybrid - ZFSCJ-2-1 [32]. The isolation of the hybrid is almost bigger than 30 dB, so the common-mode error is extremely very weak and it is ignored. Four 6 dB attenuators between the hybrid and the DUT absorb the reflection from the port. The temperature is also stable at 25 degrees

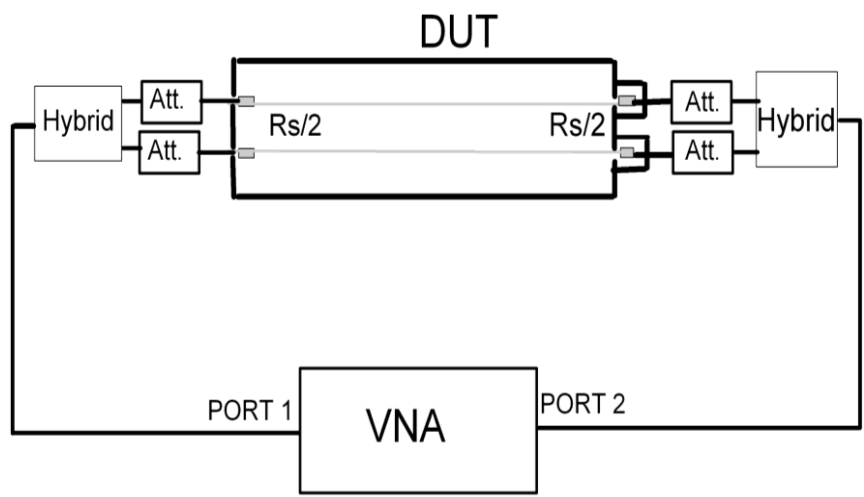

Fig 8: The schematic view of the twin-wire transverse impedance measurement

The transverse coupling impedance of the kicker can be calculated from the measured scatter coefficients $S_{21,\ DUT}$ and $S_{21,\ REF}$ as

$$Z = -2\frac{c}{\omega(2d)^2} R_c \ln(\frac{S_{21,DUT}}{S_{21,REF}}), \qquad (4)$$

with the angle frequency $\omega$.

The measured vertical coupling impedance is given in Fig 9. Due to big error below 10 MHz, the vertical impedance is shown from 10 MHz to 100 MHz. It is clear to say that there is an impedance peak in the left figure of the naked kicker on about 18 MHz as the result of the window-frame ferrite geometry and the end plate based on the CST simulation, but it disappears in the measurement of the kicker system in right figure. Moreover, the oscillation of the vertical impedance appears for the kicker system in the right picture. The oscillation is similar to the one of the longitudinal impedance, and its spacing is also 0.72 MHz, which also comes from the weak mismatch among the kicker, the PFN and the termination.

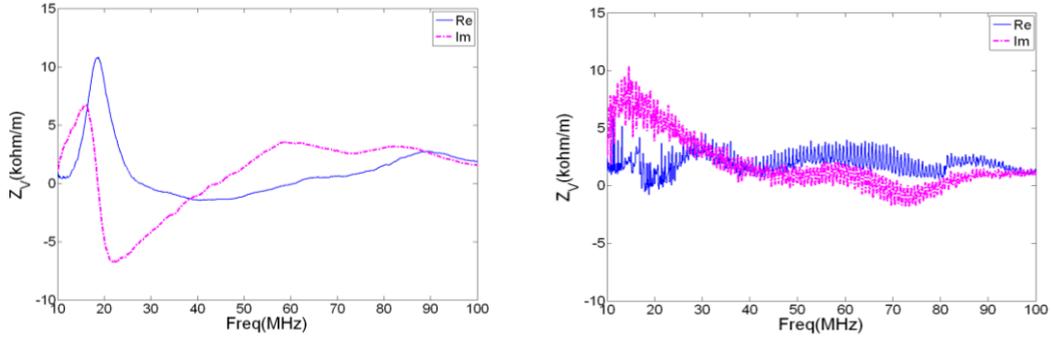

Fig 9: The measured transverse coupling impedance of the naked kicker (left) and the kicker system (right) by the twin-wire

The transverse impedance below 10 MHz by the twin-wire is not good since the error is serious. To extend the measured frequency, one turn loop is adapted to measure transverse impedance on low frequency. The schematic view of the loop method is shown in Fig 10, and the input impedances of the DUT and the REF are also measured, so the transverse impedance can be expressed as [33]

$$Z_T = \frac{c}{\omega} \frac{Z^{DUT} - Z^{REF}}{(2d)^2}, \qquad (5)$$

with the input impedance of DUT $Z^{DUT}$ and REF $Z^{REF}$.

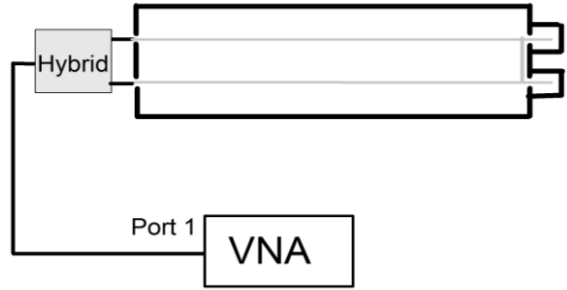

Fig 10: The schematic view of the transverse impedance measurement by the loop

The horizontal and vertical impedances of the naked kicker are measured by the loop. The horizontal coupling impedances of the naked kicker and the kicker system are shown in Fig 11, the real part of the horizontal impedance of the kicker system is almost similar to the one of the naked kicker, and the imagine parts are thinly difference. Therefore, the horizontal coupling impedance is mostly not affected by the cable, the PFN and the termination.

The vertical impedance of the naked kicker and the kicker system is shown in Fig 12. The peak on about 18 MHz in the left figure also appears for the naked kicker, which agrees well with the result of the twin-wire in Fig 9, and it also disappears for the kicker system in the right figure. The oscillation in the right figure, which comes from the reflection of the mismatch among the cable, the PFN and the termination, is excited when the kicker connects on the long cable, the PFN and the termination, and the spacing of the oscillation sideband is also 0.72 MHz, which is consistent with the result of Eq. (2).

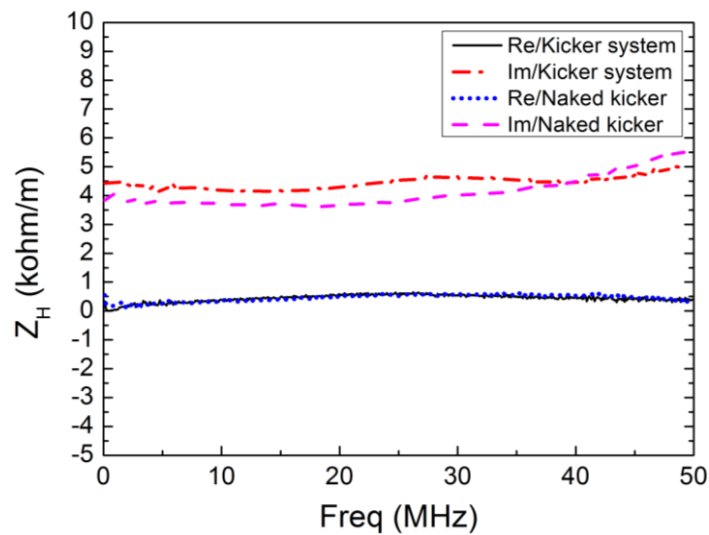

Fig 11: The measured horizontal impedance of the kicker by loop

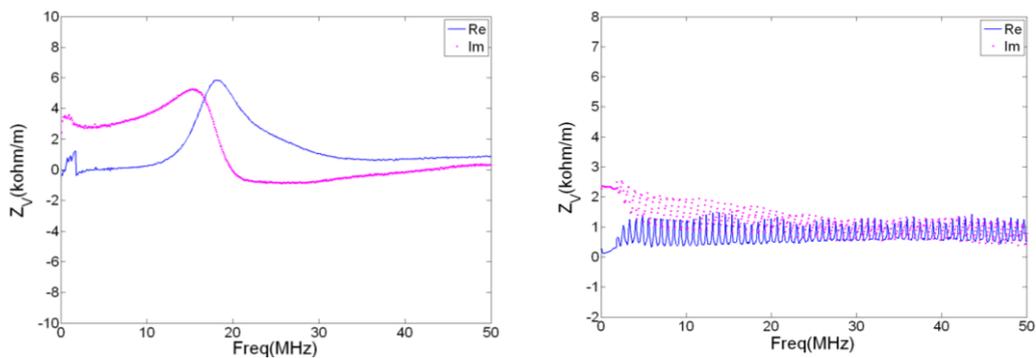

Fig 12: The measured vertical impedance of the naked kicker (left) and the kicker system (right) by loop

To find the start point of the oscillation, an improve method of the loop in Fig 13 is applied. In order to increase the signal, only the kicker with the long cable (cable open) is measured, and the vertical impedance is shown in Fig 14. It is easy to show that the start point of the oscillation is about 0.35 MHz (Owing to common-mode error, the value of measured impedance may be incorrect, it is not serious, the point of the improve method only confirms the start frequency of the oscillation). The oscillation of the mismatch from the PFN system theoretically is expressed as

$$f_{start} = \frac{c}{4\alpha L}. \tag{6}$$

The start point in Eq. (6) is 0.36 MHz, which agrees well with the measured result. Therefore, it is certain that the oscillation of the vertical impedance of the kicker system comes from the mismatch among the cable, the PFN and the termination.

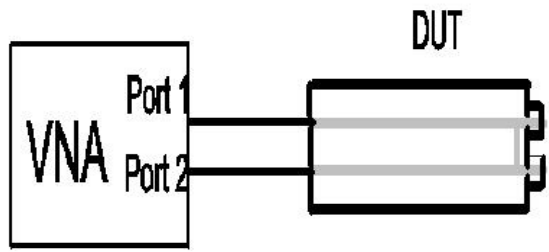

Fig 13: The improve loop method of transverse impedance measurement

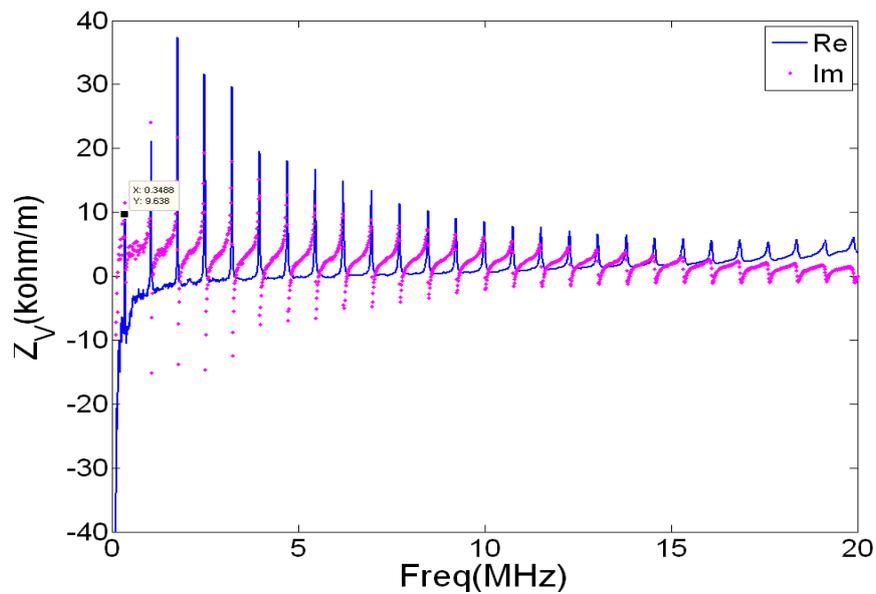

Fig 14: The measured vertical impedance of the naked kicker with the cable open by the improved loop

The transverse impedance of the naked kicker is also simulated by CST PARTICLE STUDIO, and the vertical and horizontal results of the simulation and the measurement are shown in Fig 15. It is easy to show that the difference between measurement and simulation is also small.

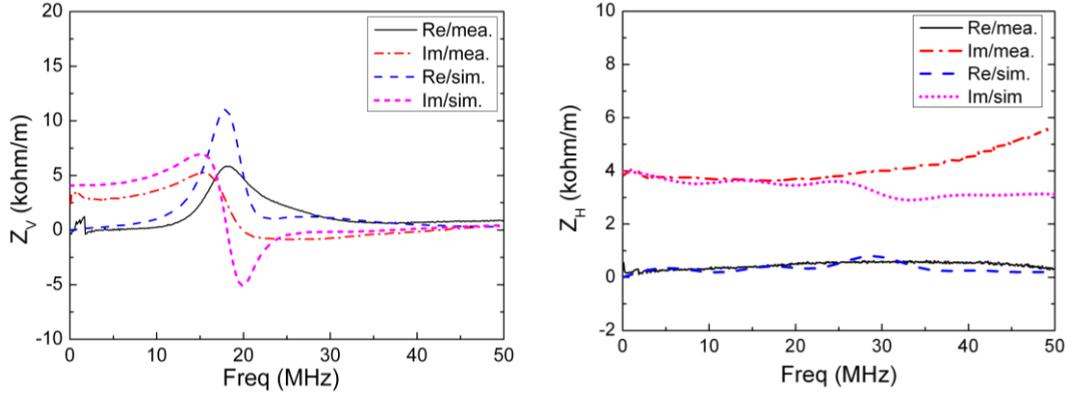

Fig 15: The simulation and measurement transverse impedance of naked kicker

The type of 8C12 ferrite ring is also adopted to restrict the oscillation of the vertical impedance. The vertical impedances of the kicker system with and without the ferrite ring are measured by the twin-wire. The oscillation is also extremely decreased in Fig 16. Therefore, the ferrite ring is also useful to absorb the reflection of the vertical impedance.

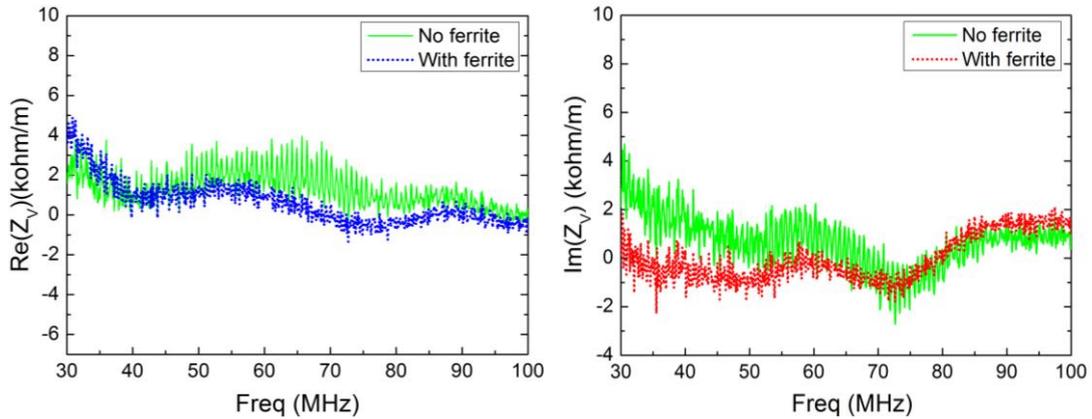

Fig 16: The vertical impedance of the kicker system with and without the ferrite ring

The transverse impedance of the others kicker can be expressed [17] as

$$Z_{\perp i} = Z_{\perp 2} \frac{(h_2)^2}{(h_i)^2}, \quad i=1,3,...8. \tag{7}$$

Here, $Z_{\perp i}$ is the transverse impedance of $i^{th}$ kicker, and $h_i$ is its height. $Z_{\perp 2}$ and $h_2$ are the measured impedance and the height of the K2, respectively. The transverse impedance of eight kicker systems are shown in Fig 17, the total vertical impedance is about $5+j10$ k$\Omega$/m and the horizontal impedance totally is about $3+j22$ k$\Omega$/m.

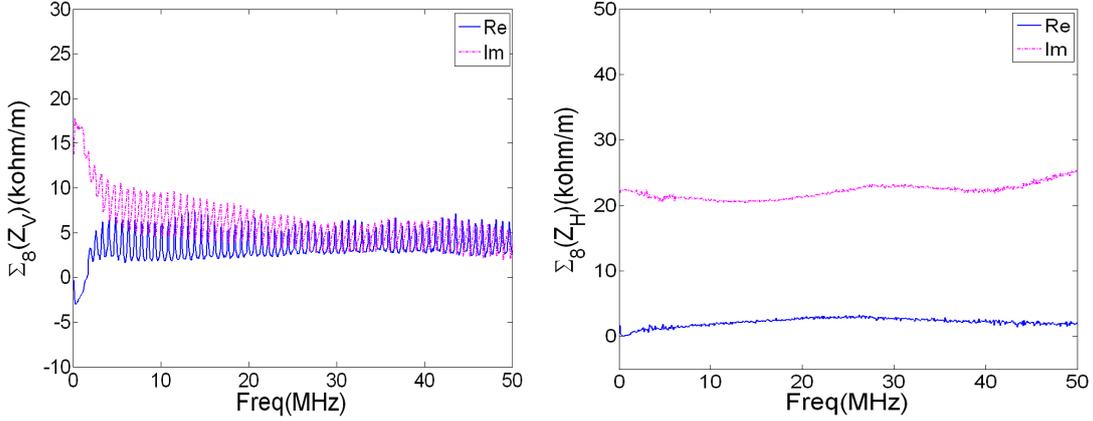

Fig 17: The total transverse impedance of the eight kickers

## 5. Summary

The CSNS/RCS includes eight modules of the ferrite kicker magnets with the window-frame geometry. The longitudinal and transverse coupling impedances of the K2 fast extraction kicker are measured by the coaxial-wire, the twin-wire and the loop method. The measured impedances are explained by comparison with the simulation of CST PARTICLE STUDIO, and they agree well. The simulation and the measurement indicate that the window-frame ferrite geometry and the end plate are the important structures causing the coupling impedance. The mismatching among the cable, the termination and the PFN affects the longitudinal and vertical coupling impedance, and 8C12 ferrite ring is useful to absorb the reflection of the mismatching. Based on the impedance measurement and simulation of the K2 kicker, the total coupling impedance of the eight kicker systems is confirmed by the scaling law. The longitudinal average impedance of the eight kicker systems is about $53+j45$ $\Omega$, the total vertical and horizontal impedance are $5+j10$ k$\Omega$/m and $3+j22$ k$\Omega$/m, respectively.

## Acknowledgements

We would like to acknowledge the support of many colleagues, especially helpful assistance from prof. Li Shen, prof. Hong Sun, Jun Zhai, Lei Wang, Jiyuan Zhai, Lin Liu and Ahong Li. The author would also like to thank Prof. Yoshiro Irie and Prof. Fritz Caspers for many discussions and comments in the measurement.